\documentclass[12pt]{article}
\usepackage[dvips]{graphicx}
\usepackage{latexsym}
\usepackage{amssymb}

\newcommand{\CP}{\mathbb{CP}}

\newcommand{\R}{\mathbb{R}}

\renewcommand{\d}{\mathrm{d}}

\newcommand{\koniec}{\hfill  $\Box $\medskip}
\def\be{\begin{equation}}
\def\ee{\end{equation}}

\def\Sm{\Sigma}

\def\om{\omega}
\def\ov{\overline}
\def\g{\mathfrak{g}}

\def\p{\partial}
\def\ov{\overline}

\def\a{\alpha}
\def\ll{\lambda}

\newtheorem{theo}{Theorem}[section]

\begin{document}
\pagestyle{plain}

\title{\vskip -70pt
\begin{flushright}
{\normalsize DAMTP-2005-26} \\
\end{flushright}
\vskip 80pt
{\bf A dispersionless integrable system associated to Diff$(S^1)$
gauge theory}\vskip 20pt}

\author{Maciej Dunajski\thanks{email M.Dunajski@damtp.cam.ac.uk}\\[15pt]
{\sl Department of Applied Mathematics and Theoretical Physics} \\[5pt]
{\sl University of Cambridge} \\[5pt]
{\sl Wilberforce Road, Cambridge CB3 0WA, UK} \\[15pt]
and \\[10pt]
George Sparling\thanks{email sparling@twistor.org} \\[15pt]
{\sl Department of Mathematics}\\[5pt]
{\sl University of Pittsburgh}\\[5pt]
{\sl Pittsburgh, PA 15260, USA}\\[15pt]
}
\date{} 
\maketitle
\begin{abstract}
A dispersionless integrable system underlying $2+1$-dimensional 
hyperCR Einstein--Weyl 
structures is obtained as a symmetry reduction of the anti--self--dual Yang--Mills equations with the gauge group Diff$(S^1)$. Two special classes of solutions
are obtained from well known soliton equations by embedding 
$SU(1,1)$ in $\mbox{Diff}(S^1)$.
\end{abstract}
\newpage
\section{From ASDYM equations to Einstein--Weyl structures}
\setcounter{equation}{0}
The idea of allowing 
infinite--dimensional groups of diffeomorphisms of some manifold $\Sm$
as gauge groups provides a link between 
the Yang--Mills--Higgs  theories 
on  $\R^n$ and conformal gravity theories on 
$\R^{n}\times\Sm$. The gauge--theoretic covariant derivatives
and Higgs fields are reinterpreted as a frame of vector fields
thus leading to a conformal structure \cite{Wa90}.  
This program has lead, among other things,  to a dual description of certain
two--dimensional  integrable systems: as symmetry reductions of 
anti--self--dual Yang--Mills (ASDYM), or as special curved anti--self--dual 
conformal structures \cite{Wa92, MW96, DMW98, D02}.

In this paper we shall give the first example of a dispersionless 
integrable system in 2+1 dimension which fits into this framework 
(Theorem \ref{asdred}).
As a spin--off we shall obtain a gauge--theoretic 
characterisation of hyperCR Einstein--Weyl spaces in 2+1 dimensions
(Theorem \ref{asdth}). We shall also construct two explicit new 
classes  of solutions to the system (\ref{PMA}) out of solutions to 
the nonlinear Schr\"{o}dinger equation, and  
the  Korteweg de Vries equation (formulae (\ref{EWnls}) and (\ref{EWkdv})).

Consider a pair of quasi-linear PDEs 
\be
\label{PMA}
u_t+w_y+uw_x-wu_x=0,\qquad u_y+w_x=0,
\ee
for  two real functions $u=u(x, y, t), w=w(x, y, t)$.
This integrable system has recently been used to characterise a class
of Einstein--Weyl structures in 2+1 dimensions \cite{D04}. It has also
appeared in other contexts \cite{Pa03, MSh03, MSh04, FK04} as an example of 2+1 dimensional dispersionless integrable models.

The equations (\ref{PMA}) arise as  compatibility conditions
$[L, M]=0$ of an overdetermined system of linear equations
$L\Psi=M\Psi=0$, where $\Psi=\Psi(x, y, t, \ll)$ is a function, 
$\ll$ is a spectral parameter, and the Lax pair is given by
\be
\label{Lax11}
L=\p_t-w\p_x-\ll\p_y,\qquad
M=\p_y+u\p_x-\ll\p_x.
\ee
This should be contrasted with Lax pairs for other
dispersionless integrable systems \cite{Ta90, Z94, GMM, Kon3, BK04}
which contain derivatives w.r.t the spectral parameter.

The first equation in (\ref{PMA}) resembles a flatness
condition for a connection with the underlying Lie algebra diff$(\Sm)$,
where $\Sm=S^1$ or $\R$. The following result makes this 
interpretation precise  
\begin{theo}
\label{asdred}
The system {\em(\ref{PMA})} arises as a symmetry reduction 
of the anti--self--dual Yang Mills  equations in signature $(2, 2)$ 
with the infinite--dimensional gauge
group Diff$(\Sm)$ and two commuting translational symmetries 
exactly one of which is 
null. Any such symmetry reduction  is gauge equivalent to
(\ref{PMA}).
\end{theo}
{\bf Proof.}
Consider the flat metric of signature $(2, 2)$ on $\R^4$ which
in double null coordinates $y^{\mu}=(t, z, \tilde{t}, \tilde{z})$ takes the 
form 
\[
\d s^2=\d t\d \tilde{t}-\d z\d \tilde{z},
\]
and choose the volume element 
$\d t\wedge\d \tilde{t}\wedge\d z\wedge\d \tilde{z}$.
Let $A\in T^*\R^4\otimes\g$ be a connection one--form, and
let $F$ be its curvature two--form. Here $\g$ is the Lie algebra of some
(possibly infinite dimensional) gauge group $G$.
In a local trivialisation $A=A_\mu\d y^\mu$
and $F=(1/2)F_{\mu\nu}\d y^{\mu}\wedge\d y^{\nu}$, where
$F_{\mu\nu}=[D_{\mu}, D_{\nu}]$
takes its values in $\g$.  Here
$D_\mu=\p_\mu-A_\mu$ is the covariant derivative.
The connection is defined up to gauge transformations
$A\rightarrow b^{-1}Ab-b^{-1}\d b$, where $b\in \mbox{Map}(\R^4, G)$.
The ASDYM equations on $A_\mu$ are $F=-\ast F$, or 
\[
F_{tz}=0, \qquad F_{t\tilde{t}}-F_{z\tilde{z}}=0,\qquad 
F_{\tilde{t}\tilde{z}}=0.
\]
These equations are equivalent to the commutativity of the Lax pair
\[
L=D_t-\ll D_{\tilde{z}}, \qquad M=D_z-\ll D_{\tilde{t}}
\]
for every value of the parameter $\ll$.

We shall  require that the connection possesses two commuting
translational symmetries, one null and one non--null
which in our coordinates are in $\p_{\tilde{t}}$ 
and $\p_{\tilde{y}}$ directions,
where $z=y+\tilde{y}, \tilde{z}=y-\tilde{y}$.
Choose a gauge such that $A_{\tilde{z}}=0$ and one of the 
Higgs fields $\Phi=A_{\tilde{t}}$ is  constant. The Lax pair has so
far been reduced to
\be
\label{2Dlax}
L=\p_t-W-\ll\p_y, \qquad M=\p_y-U-\ll\Phi,
\ee
where $W=A_t$ and $U=A_z$ are 
functions of $(y, t)$ with values in the Lie algebra 
$\g$, and $\Phi$ is an element of $\g$ which doesn't depend on $(y, t)$.
The reduced ASDYM equations are
\[
\p_yW-\p_tU+[W, U]=0, \qquad \p_y U+[W, \Phi]=0.
\]
Now choose $G=\mbox{Diff}(\Sigma)$, where $\Sm$ is some one--dimensional 
manifold, so that $(U, W, \Phi)$ become vector fields
on $\Sigma$. We can choose a local
coordinate $x$  on $\Sigma$ such that 
\be
\label{from1}
\Phi=\p_x, \qquad W=w(x, y, t)\p_x, \qquad  U=-u(x, y, t)\p_x
\ee
where $u, w$ are smooth functions on $\R^3$. The reduced
Lax pair (\ref{2Dlax}) is identical to (\ref{Lax11}) and the ASDYM equations
reduce to the pair of PDEs (\ref{PMA}). \koniec

Recall that a Weyl structure on an $n$ dimensional manifold ${\cal W}$ consists
of  a torsion-free connection $D$ and a conformal structure 
$[h]$ which is compatible with $D$ in a sense
that $Dh=\om\otimes h$ for some one-form $\om$ and $h\in[h]$. 
We say that a Weyl structure
is Einstein--Weyl if the traceless part of the symmetrised Ricci tensor of
$D$ vanishes. The three--dimensional Einstein--Weyl structure is called hyperCR \cite{CP99,
GT98, DT01, D04} if
its mini-twistor space \cite{H82} is a holomorphic bundle over $\CP^1$.

In \cite{D04} it was demonstrated that if $n=3$, and $[h]$ has signature 
$(++-)$ then 
all Lorentzian hyperCR Einstein--Weyl structures are locally of the form 
\be
\label{PMAEW}
h=(\d y+u\d t)^2-4(\d x+w\d t)\d t,\qquad 
\om =u_x\d y+(uu_x+2u_y)\d t,
\ee
where $u, w$ satisfy (\ref{PMA}).
This result combined with Theorem \ref{asdred} yields the following
coordinate independent characterisation of the hyperCR Einstein--Weyl
condition
\begin{theo}
\label{asdth}
The ASDYM equations 
in $2+2$ dimensions with two commuting translational symmetries
one null and one non--null, and the gauge group Diff$(\Sm)$
are gauge equivalent to the hyperCR Einstein--Weyl equations in 
\em{2+1} dimensions. 
\end{theo}
This is a Lorentzian analogue of a Theorem
proved in \cite{C01} in the Euclidean case. The readers should note
that in \cite{C01} the result is formulated in terms of the Hitchin system,
and not reductions of the ASDYM system.
\section{Reductions to KdV and NLS}
\setcounter{equation}{0}
Reductions of the ASDYM equations with $G=SU(1, 1)$
by two translations (one of which is null)
lead to well--known integrable systems 
KdV, and NLS \cite{MS89}. The group $SU(1, 1)$ is a subgroup
of Diff$(S^1)$ which can be seen by considering the Mobius 
action of $SU(1, 1)$
\[
\zeta\longrightarrow M(\zeta)=\frac{\a\zeta+\beta}{\ov{\beta}\zeta+\ov{\a}}, \qquad
|\a|^2-|\beta|^2=1
\]
on the unit disc. This restricts to the action on the circle as
$|M(\zeta)|=1$ if $|\zeta|=1$.
We should therefore expect that  equation (\ref{PMA}) contains
KdV and NLS as its special cases (but not necessarily symmetry
reduction). To find  explicit classes of solutions to (\ref{PMA}) out of solutions 
to KdV and NLS we proceed as follows.
Consider the matrices
\[
\tau_+=\left(
\begin{array}{cc}
0&1\\
0&0
\end{array}
\right ), \qquad
\tau_-=\left(
\begin{array}{cc}
0&0\\
1&0
\end{array}
\right ),\qquad 
\tau_0=\left(
\begin{array}{cc}
1&0\\
0&-1
\end{array}
\right )
\]
with the commutation relations
\[
[\tau_+, \tau_-]=\tau_0, \qquad [\tau_0, \tau_+]=2\tau_+, \qquad
[\tau_0, \tau_-]=-2\tau_-.
\]
The NLS equation 
\be
\label{nls}
i\phi_t=-\frac{1}{2}\phi_{yy}+\phi|\phi|^2, \qquad \phi=\phi(y, t)
\ee
arises from the reduced Lax pair (\ref{2Dlax}) with
\[
W=\frac{1}{2i}(-|\phi|^2\tau_0+\phi_y\tau_--\ov{\phi}_y\tau_+),
\qquad 
U=-\phi\tau_--\ov{\phi}\tau_+,
\qquad 
\Phi=i\tau_0.
\]
Now we replace the matrices by vector fields on $\Sm$ corresponding 
to the embedding of $su(1, 1)$ in diff$(\Sm)$ 
\[
\tau_+\longrightarrow \frac{1}{2i}e^{2ix}\frac{\p}{\p x}, \qquad
\tau_-\longrightarrow -\frac{1}{2i}e^{-2ix}\frac{\p}{\p x}, \qquad
\tau_0\longrightarrow \frac{1}{i}\frac{\p}{\p x},
\]
and read off the solution to (\ref{PMA}) from (\ref{from1})
\be
\label{EWnls}
u=\frac{1}{2i}(\ov{\phi}e^{2ix}-\phi e^{-2ix}), \qquad
w=\frac{1}{2}|\phi|^2+\frac{1}{4}(e^{2ix}\ov{\phi}_y+e^{-2ix}\phi_y).
\ee
The second equation in (\ref{PMA}) is satisfied identically, and the
first is satisfied if $\phi(y, t)$ 
is a solution to the NLS equation (\ref{nls}).

Analogous procedure can be applied to the KdV
equation
\be
\label{KdV}
4v_t-v_{yyy}-6vv_y=0, \qquad v=v(y, t).
\ee
The Lax pair for this equation
is given by (\ref{2Dlax}) with
\[
W=\Big(q_y\tau_+
-\kappa\tau_- -\Big(\frac{1}{2}q_{yy}+qq_y\Big)\tau_0\Big),
\qquad 
U=\tau_+-q\tau_0 -(q_y+q^2)\tau_-,
\qquad \Phi=\tau_-,
\]
where 
\[
\kappa=\frac{1}{4}q_{yyy}+qq_{yy}+\frac{1}{2}{q_y}^2+q^2q_y,\qquad
\mbox{and}\; v=2q_y.
\]
Now we choose $x$ such that
\[
\tau_+\longrightarrow -x^2\frac{\p}{\p x}, \qquad
\tau_-\longrightarrow \frac{\p}{\p x}, \qquad
\tau_0\longrightarrow 2x\frac{\p}{\p x},
\]
and read off the
expressions for $u$ and $w$
\be
\label{EWkdv}
u=x^2+2xq+q_y+q^2, \qquad w=-x^2q_y-x(q_{yy}+2qq_y)-\kappa.
\label{KdVsol}
\ee
The second equation in (\ref{PMA}) holds identically, and the
first is satisfied if $v$ is a solution to (\ref{KdV}). 

In references \cite{MSh03, MSh04} the so called `universal hierarchy' was 
studied and a general procedure of constructing its 
differential reductions was proposed. 
The system (\ref{PMA}) arises from the first two flows of this hierarchy, 
but it is not clear how the differential constraints 
imposed in \cite{MSh03, MSh04} can be understood from the Diff$(S^1)$ point of 
view. It would be interesting to see whether our reductions to NLS and KdV
are `differential' in the sense of the above references.

One remark is in place: There is a standard procedure \cite{JT85} of 
constructing
anti--self--dual conformal structures with symmetries out of EW
structures in 3 or 2+1 dimensions. The procedure is based on
solving a linear 
generalised monopole equation on the EW background.
Moreover, the hyperCR EW structures 
always lead to hyper--complex conformal structures with a 
tri--holomorphic Killing vector, and it is possible to 
choose a monopole such that there exist a Ricci--flat metric in the conformal 
class \cite{GT98}. Any  hyperCR EW (\ref{PMAEW})
structure given in terms of KdV, or NlS potential by (\ref{EWkdv}) or 
(\ref{EWnls}) will therefore lead to a $(++--)$ ASD Ricci--flat metric
with a tri--holomorphic homothety. 
The explicit formulae for the metric in terms 
of solutions to (\ref{PMA}) can be found in \cite{D04}.
Another class of ASD Ricci--flat metrics  has been constructed from 
KdV and NLS, by embedding $SU(1,1)$ in a Lie algebra of volume preserving
transformations of the  Poincare disc \cite{DMW98}. These metrics generically 
do not admit any symmetries, and therefore are different from ours.
\section*{Acknowledgements}
Both authors 
were partly supported by NATO grant PST.CLG.978984. MD is a member of the
European Network in Geometry, 
Mathematical Physics and Applications. We wish to thank 
the anonymous referee for valuable comments.

\end{document}